\documentclass[12pt,a4paper,draft]{amsart}
\usepackage{amsmath,amscd,amssymb}
\newcommand{\Ii}{{\mathrm i}}
\newcommand{\E}{{\mathrm e}}
\newcommand{\cprime}{\/{\mathsurround=0pt$'$}}
\newcommand{\CC}{\mathcal{C}}
\newcommand{\CE}{\mathcal{E}}
\newcommand{\CF}{\mathcal{F}}
\newcommand{\CL}{\mathcal{L}}
\newcommand{\CO}{\mathcal{O}}
\newcommand{\CU}{\mathcal{U}}
\newcommand{\Dr}{\mathrm{D}}
\newcommand{\CDr}{\mathop{\CC\mathrm{D}}}
\newcommand{\Der}{\mathop{\mathrm{Der}}}
\newcommand{\End}{\mathop{\mathrm{End}}}
\newcommand{\Aut}{\mathop{\mathrm{Aut}}}
\newcommand{\ad}{\mathop{\mathrm{ad}}\nolimits}
\newcommand{\SL}{\mathop{\mathrm{SL}_2}}
\newcommand{\Sl}{\mathop{\mathrm{sl}_2}}

\newtheorem{theorem}{Theorem}[section]

\newtheorem{proposition}[theorem]{Proposition}

\theoremstyle{definition}
\newtheorem{definition}{Definition}[section]
\newtheorem{example}{Example}[section]

\theoremstyle{remark}
\newtheorem{remark}{Remark}[section]

\def\ldb{{\rm [\![}}
\def\rdb{{\rm ]\!]}}
\newcommand{\fnij}[2]{\ldb{#1},{#2}\rdb}

\advance\textheight-2ex
\advance\footskip 2ex
\begin{document}
\title{Coverings and Integrability\\
of the Gauss--Mainardi--Codazzi Equations}
\author{Joseph Krasil{\cprime}shchik, Michal Marvan}

%\Year{1998}
%\Date{December 4}
%\Number{8/98}
%\MakeTitle
%\author{\tauthor}

\thanks{The first author was partially supported by RFBR grant
97-01-00462 and INTAS grant 96-0793.
This work was finished during author's stay at ESI in October 1998.}
\address{Moscow Institute for Municipal Economy and
Independent University of Moscow\protect\newline
Correspondence to:
1st Tverskoy-Yamskoy per. 14, Apt. 45, 125047 Moscow, Russia}
\email{josephk@glasnet.ru}

\address{Dept. of Mathematics, Silesian University at Opava,
Bezru\v{c}ovo n\'am. 13, 746 01 Opava, Czech Republic}
\email{Michal.Marvan@fpf.slu.cz}
\thanks{The second author was supported from Project VS~96003
(``Global Analysis'') of the
Ministry of Education, Youth and Sports, Czech Republic, and grant
No.~201/98/0853 from the Czech Grant Agency}


\begin{abstract}
Using covering theory approach (zero-curvature representations with the
gauge group $\SL$), we insert the spectral parameter into the
Gauss--Mainardi--Codazzi equations in Tchebycheff and geodesic coordinates.
For each choice, four integrable systems are obtained.
\end{abstract}


\maketitle

\section*{Introduction}
When immersed in the Euclidean space $E^3$, surfaces with
metric $g = g_{11}\,dx^2 + 2 g_{12}\,dx\,dy + g_{22}\,dy^2$
and the second fundamental form
$b = b_{11}\,dx^2 + 2 b_{12}\,dx\,dy + b_{22}\,dy^2$
satisfy the Gauss--Mainardi--Codazzi equations (GMC)
$R^k_{ijl} = b_{ij} b^k_l - b_{il} b^k_j$ and $b_{ij,k} = b_{ik,j}$,
where $R^k_{ijl}$ are components of the curvature tensor.
By imposing an algebraic constraint of the form
$L(g_{ij},b_{ij},x,y) = 0$ we obtain a special instance of what is
called \emph{reduced GMC equations}.
Many reduced GMC systems have been found integrable in the sense of
soliton theory, e.g., in works \cite{B-N,B1,B2,C,C-G-S,L-S,M-S}, thus
leading to \emph{integrable} classes of surfaces.
An example is provided by the so-called \emph{linear Weingarten surfaces}
determined by a linear relation $\alpha K + \beta H = \gamma$
($\alpha,\beta,\gamma = \mathrm{const}$) between their Gauss and main
curvatures, see the recent book \cite{Te} and references therein.

In this paper we apply the methods of~\cite{M} to a related problem.
Namely, we check GMC equations, written here in Tchebycheff and geodesic
coordinates, for existence of a zero-curvature representation (ZCR) with
coefficients in the complex Lie algebra $\Sl$.
For each choice of the coordinates we have found four cases possessing a
non-removable parameter.

An \emph{$\Sl$-valued ZCR} is determined by an $\Sl$-valued form
$A\,dx + B\,dy$ satisfying $A_{,y} - B_{,x} + [A,B] = 0$.
A \emph{gauge-equivalent} ZCR is given by $A'\,dx + B'\,dy$ with
$A' = S_{,x} S^{-1} + S A S^{-1}$, $B' = S_{,y} S^{-1} + S B S^{-1}$
for a suitable $\SL$-matrix~$S$.
Any ZCR gauge-equivalent to the zero form is called \emph{trivial}.
A non-removable parameter is a parameter that cannot be removed by a
gauge transformation.

As it happens, the theory of zero-curvature representations is naturally
formulated in terms of coverings \cite{K-V}. To obtain a well-defined
construction, one needs to consider a linear covering \cite{Ts} endowed with
an action of a Lie group $G$. Then a ZCR related to this covering is
determined by a closed $\mathfrak{g}$-valued horizontal $1$-form. To make
exposition self-contained, we expose here a geometrical theory of ZCR based
on the covering theory in the category of differential equations.

\section{Linear coverings and zero-curvature representations}

Let $\CO$ be a smooth manifold (possibly, infinite-dimensional) with
an integrable finite-dimensional distribution $\CC$ of dimension
$n$. Integrability is understood in the formal sense here: a distribution is
said to be integrable, if the module of vector fields lying in this
distribution is closed with respect to commutator. Recall (see \cite{K-V})
that a \emph{covering} over the pair $(\CO,\CC)$ is a locally trivial fiber
bundle $\tau\colon\tilde{\CO}\to\CO$ such that
\begin{enumerate}
\item The manifold $\tilde\CO$ is endowed with an $n$-dimensional
integrable distribution $\tilde\CC$.
\item The mapping $d\tau\big|_{\tilde{\CC}_{\tilde{\theta}}}\colon
\tilde{\CC}_{\tilde{\theta}}\to T_\theta\CO$ is an isomorphism onto the
plane $\CC_{\tau(\tilde{\theta})}\subset T_\theta\CO$ for any
$\tilde{\theta}\in\tilde{\CO}$.
\end{enumerate}
Our main concern will be with manifolds $\CO$ of the form $\CE^\infty$,
where $\CE^\infty$ is the infinite prolongation of a differential equation
$\CE\subset J^k(\pi)$, $J^k(\pi)$ being the manifold of $k$-jets for some
locally trivial bundle $\pi\colon E\to M$. The corresponding distribution
$\CC$ is the \emph{Cartan distribution} (see details in \cite{K-L-V}).

Denote by $\Dr(N)$ the Lie algebra of vector fields on a smooth manifold
$N$ and by $\CDr(\CO)\subset\Dr(\CO)$ the subalgebra of vector fields lying
in the distribution $\CC$. Then a covering structure in the bundle $\tau
\colon\tilde{\CO}\to\CO$ is identified with a flat \emph{$\CC$-connection}
in $\tau$, i.e., with a mapping $\nabla=\nabla^\tau\colon\CDr(\CO)\to
\Dr(\tilde{\CO})$ such that
\begin{enumerate}
\item $\nabla$ is an $\CF(\CO)$-linear mapping, where $\CF(\CO)$ is the
algebra of smooth functions on $\CO$.
\item For any $X\in\CDr(\CO)$, the field $\nabla_X\in\Dr(\tilde\CO)$ projects
to $X$ by $d\tau$.
\item For any $X$, $Y\in\CDr(\CO)$ one has
\begin{equation}
\label{eq:conn}
\nabla_{[X,Y]}=[\nabla_X,\nabla_Y].
\end{equation}
\end{enumerate}
We shall also say that $\nabla$ determines a \emph{covering structure} in
$\tau$.

Two coverings $\tau_1\colon\tilde{\CO}_1\to\CO$, $\tau_2\colon\tilde{\CO}_2
\to\CO$ with the connections $\nabla^1$ and $\nabla^2$ respectively are said
to be \emph{equivalent}, if there exists an equivalence $\varphi$ of the
bundles $\tau_1$, $\tau_2$, i.e., a
diffeomorphism $\varphi\colon\tilde{\CO}_1\to\tilde{\CO}_2$ satisfying
$\tau_1=\tau_2\circ\varphi$, such that $d\varphi\circ\nabla^1=\nabla^2$.

If $\CU\subset\CO$ is a trivialization of the bundle $\tau$, i.e., a domain
such that the bundle $\tau\big|_{\tau^{-1}(\CU)}\colon\tau^{-1}(\CU)\to\CU$
is trivial, then the covering structure over $\CU$ is given by the splitting
\begin{equation}
\label{eq:split}
\nabla_X=X+V_X,\quad V_X\in\Dr(\tau^{-1}(\CU)),
\end{equation}
where $V_X$ is a $\tau$-vertical vector field, while the flatness condition
is expressed in the form
\begin{equation}
\label{eq:flat}
V_{[X,Y]}=[X,V_Y]+[V_X,Y]+[V_X,V_Y].
\end{equation}

Let now $\tau$ be a vector bundle and $\CL\CF(\tilde\CO)$ denote the
$\CF(\CO)$-module of fiber-wise linear smooth functions on $\tilde\CO$.

\begin{definition}
\label{def:lin}
Let $\tau\colon\tilde{\CO}\to\CO$ be a vector bundle.
\begin{enumerate}
\item
A covering $\tau\colon\tilde{\CO}\to\CO$ is called
\emph{linear}, if the field $\nabla_X$ preserves the module
$\CL\CF(\tilde{\CO})$ for any $X\in\CDr(\CO)$ (cf.~\cite{Ts}, where the
same concept is introduced, though without using the term ``covering'').
\item Two linear coverings are said to be \emph{equivalent}, if there exists
their equivalence $\varphi$, which is a morphism of vector bundles (i.e.,
is fiber-wise linear).
\end{enumerate}
\end{definition}

Let $\CU\subset\CO$ be a trivialization of the bundle $\tau$ and $w^1,\dots,
w^r,\dots$ be coordinates along the fiber. Then a covering $\tau$ is linear
if and only if the field $V_X$ represents in the form
\begin{equation}
\label{eq:loc}
V_X=\sum_{r}\Big(\sum_{s}V_X^{rs}w^s\Big)\frac{\partial}{\partial w^r}
\end{equation}
for any $X\in\CDr(\CO)$, where $V_X^{rs}$ are smooth functions on $\CO$ and
all internal sums are finite. Thus, any linear covering is locally
determined by the system of matrices $\Vmatrix V_i^{rs}\endVmatrix$, where
$V_i^{rs}=V_{X_i}^{rs}$, $X_i$, $i=1,\dots,n$, being a local basis of the
distribution $\CC$.

If an automorphism of the vector bundle $\tau$ is locally given by a matrix
$S$, then the matrices $V_X$ are transformed by the formula
\begin{equation}
\label{eq:trans}
V_X\mapsto X(S)S^{-1}+SV_XS^{-1},
\end{equation}
where $X(S)$ denotes the component-wise action.

\begin{example}
\label{exmp:lin*}
Let $\tau\colon\tilde{\CO}\to\CO$ be a covering.
Denote by $\CC\Lambda^1(\CO)\subset\Lambda^1(\CO)$ the
$\CF(\CO)$-submodule consisting of forms such that $\mathrm{i}_X\omega=0$
for any $X\in\CC\Dr(\CO)$. In a similar way, we obtain the
$\CF(\tilde{\CO})$-submodule $\tilde{\CC}\Lambda^1(\tilde{\CO})$ in
$\Lambda^1(\tilde{\CO})$. Then, since $\tilde{\CC}\Dr(\tilde{\CO})$ projects
to $\CC\Dr(\CO)$ by $\tau_*$, one has $\tau^*(\CC\Lambda^1(\CO))\subset
\tilde{\CC}\Lambda^1(\tilde{\CO})$. Denote by $\CC\Lambda^1(\tilde{\CO})
\subset\tilde{\CC}\Lambda^1(\tilde{\CO})$ the $\CF(\tilde{\CO})$-submodule
generated by $\tau^*(\CC\Lambda^1(\CO))$. Then $\CC\Lambda^1(\tilde{\CO})$
is stable with respect to the Lie action of vector fields lying in
$\CC\Dr(\tilde{\CO})$ and we can extend this action to the quotient module
$\tilde{\CC}\Lambda^1(\tilde{\CO})/\CC\Lambda^1(\tilde{\CO})$. One can
easily see that the corresponding vector bundle $\tau_\ell\colon
T_\CC^*(\tilde{\CO})\to\tilde{\CO}$ is endowed in this way with a linear
covering structure.
\end{example}

\begin{example}
\label{exmp:lin}
Consider a construction dual to that of
Example~\ref{exmp:lin*}. Let again $\tau\colon\tilde{\CO}\to\CO$ be an
arbitrary covering. In the tangent bundle $T\tilde{\CO}\to\tilde\CO$, take
the subbundle $\tau^\ell\colon T^{\mathrm{v}}\tilde{\CO}\to\tilde{\CO}$ of
\emph{$\tau$-vertical} tangent vectors. Obviously, the bundle $\tau^\ell$ is
dual to $\tau_\ell$. If a vector field $Y\in\Dr(\tilde\CO)$ is such that
$d\tau Y_{\tilde\theta}$ lies in the plane $\CC_\theta$, $\theta=
\tau(\tilde{\theta})$, we can define its \emph{vertical component}
$Y^{\mathrm{v}}$ by setting
\[
Y^{\mathrm{v}}_{\tilde{\theta}}:=Y_{\tilde{\theta}}-
\nabla(d\tau(Y_{\tilde{\theta}})),
\]
at any point $\tilde{\theta}\in\tilde{\CO}$.
Take a $\tau$-vertical vector field $Z$, i.e., a section of the bundle
$\tau^\ell$, and a field $X\in\tilde{\CC}\Dr(\tilde{\CO})$.
Then the field $[X,Z]$ possesses the above formulated property and the
relation
\[
\tilde{\nabla}_X(Z):=[X,Z]^\mathrm{v}
\]
determines a covering structure in the
projection $\tau^\ell\colon T^{\mathrm{v}}\tilde{\CO}\to\tilde{\CO}$.
This covering structure in the bundle $\tau^\ell$ is linear
and is called the \emph{linearization} of the covering $\tau$
(cf.~\cite{K1,M1}).
\end{example}

\begin{example}
\label{exmp:end}
Let $A$ be a commutative $\Bbbk$-algebra, $\Bbbk$ being a
commutative ring, and $P$ be an
$A$-module. Recall that a \emph{derivation} of the module $P$ is a linear
mapping $\bar{X}\colon P\to P$ satisfying $\bar{X}(ap)=X(a)p+a\bar{X}(p)$ for
all $a\in A$, $p\in P$, and some derivation $X\colon A\to A$ of the algebra
$A$. Denote the $A$-module of all such derivations by $\Der(P)$. Obviously,
$\Der(P)$ is a Lie $\Bbbk$-algebra with respect to the commutator.

In particular, let $\tau\colon\tilde{\CO}\to\CO$ be a linear covering. Then,
by definition, the covering structure in $\tau$ determines the homomorphism
of Lie algebras $\nabla\colon\CDr(\CO)\to\Der(\CL\CF(\tilde{\CO}))$, which
splits the natural projection $\Der(\CL\CF(\tilde{\CO}))\to\Dr(\CO)$,
$\bar X\mapsto X$. The kernel of this projection coincides with the module
$\End(\CL\CF(\tilde{\CO}))=\Gamma(\End(\tau))$, where $\Gamma(\cdot)$
denotes the module of sections. This kernel is an ideal of the Lie algebra
$\Der(\CL\CF(\tilde{\CO}))$. Thus, we obtain a linear covering structure in
the bundle $\End(\tau)$.

In local coordinates, an endomorphism $\varphi$ is represented by a matrix
$A_\varphi$ and the action of a field $X\in\CDr(\CO)$ on $A_\varphi$ is
expressed by the formula
\[
X(A_\varphi)+[V_X,A_\varphi],
\]
where $X(A_\varphi)$ is understood as the component-wise action.
\end{example}

Assume now that a Lie group $G$ acts in the bundle $\tau$ by linear
automorphisms of this bundle, i.e., a representation $\rho\colon G\to
\Aut(\tau)$ is given. Let $\mathfrak{g}$ be the corresponding Lie algebra.
Then the representation $\rho$ yields the following commutative diagram:
\[
\CD
\CO\times\mathfrak{g}@>\bar{\rho}>>T^\mathrm{v}\tilde{\CO}\\
@V\mathrm{pr}_{\CO}VV                     @VV{\tau^\ell}V\\
\CO@<<{\tau}<\tilde{\CO}
\endCD
\]
\begin{definition}
\label{def:zcr}
Let $\nabla^0$, $\nabla^1$ be two linear covering
structures in the vector bundle $\tau\colon\tilde{\CO}\to\CO$.
\begin{enumerate}
\item We say that the covering structures $\nabla^0$ and $\nabla^1$ differ
by a
\emph{zero-curvature representation} (with the \emph{gauge group $G$}, or
with the \emph{gauge algebra $\mathfrak{g}$}), if for any $X\in\CDr(\CO)$
there exists a $\mathfrak{g}$-valued function $g_X\in
\Gamma(\mathrm{pr}_\CO)$ linear in $X$ and such that
\[
\nabla_X^1-\nabla_X^0=\bar\rho\circ g_X\circ\tau.
\]
In other words, the vertical vectors $\nabla_X^1$ and $\nabla_X^0$ differ by
an element of the Lie algebra $\mathfrak{g}$ at any point $\theta\in\CO$.
\item Let $\nabla^0$, $\nabla^1$ and $\square^0$, $\square^1$ be two pairs
satisfying the above definition. They are said to be \emph{equivalent}, if
there exists an element $g\in G$ such that $\rho(g)$ is an equivalence of
$\nabla^0$ to $\square^0$ and of $\nabla^1$ to $\square^1$.
\end{enumerate}
\end{definition}

Thus, equivalence in the sense of the previous definition means that the
matrices $S$ in \eqref{eq:trans} belong to the group $G$.

\begin{remark}
\label{rem:vulgar}
In coordinate computations for particular differential
eq\-uations $\CE^\infty=\CO$, the bundle $\tau$ is usually trivial, $\tau=
\mathrm{pr}_{\CE^\infty}\colon\CO=\CE^\infty\times\mathbb{R}^r\to\CE^\infty$,
and the structure $\nabla^0$ is also trivial (i.e., is such that the matrices
$V_X$ vanish for all $X\in\CDr(\CE^\infty)$, see eq.~\eqref{eq:loc}). This
should be the reason why it is never taken into consideration explicitly
and is accepted ``by default'': only the covering $\nabla^1$ is included
into definition.
\end{remark}

From now on we take $\CO=\CE^\infty$, fix a vector bundle $\tau\colon
\tilde{\CE}^\infty\to\CE^\infty$, where $\CE\subset J^k(\pi)$ is a
differential equation, and assume that a Lie group $G$ acts in $\tau$ in the
above described way. Consider a covering structure $\nabla^0$ in the bundle
$\tau$. Then $\nabla^0$ determines a flat connection in the bundle
$\tilde{\CE}^\infty\xrightarrow{\tau\ }\CE^\infty\xrightarrow{\pi_\infty\ }M$
extending the Cartan connection in $\pi_\infty\colon\CE^\infty\to M$. This
connection is uniquely determined by the corresponding connection form $U^0$
and the condition $\nabla^0$ to be a covering structure is equivalent to the
identity $\fnij{U^0}{U^0}=0$, where $\fnij{\cdot\,}{\cdot}=0$ is the
\emph{Fr\"{o}licher--Nijenhuis bracket} (see \cite{K} for more details).

If $\nabla^1$ is another covering structure in $\tau$, then the connection
forms $U^0$ and $U^1$ differ by a horizontal vector-valued form $\omega_{01}
=U^1-U^0$. From \cite{K} we immediately obtain the following
\begin{proposition}
Two linear covering structures in the vector bundle $\tau$ satisfy
Definition~\emph{\ref{def:zcr}\,(1)} if and only if
\begin{enumerate}
\item[\emph{1.}] The form $\omega_{01}$ is $\mathfrak{g}$-valued.
\item[\emph{2.}] The identity
\begin{equation}
\label{eq:CMeq}
\textstyle
\partial^0=\omega_{01}+\frac12\fnij{\omega_{01}}{\omega_{01}}=0
\end{equation}
holds, where $\partial^0=\fnij{\omega_{01}}{\cdot}$ is the differential
associated to $\nabla^0$ by the Fr\"{o}licher--Nijenhuis bracket.
\end{enumerate}
\end{proposition}
\begin{remark}
\label{rem:vulgar1}
In the situation discussed in Remark~\ref{rem:vulgar},
the action of the differential $\partial^0$ on the form $\omega_{01}$
coincides with the component-wise action of $-d_h$, where $d_h$ is the
\emph{de~Rham horizontal differential} on $\CE^\infty$. Thus,
eq.~\eqref{eq:CMeq} can be rewritten as
\[
\textstyle
d_h\omega_{01}=\frac12\fnij{\omega_{01}}{\omega_{01}}
\]
in this case.
\end{remark}

\section{Tchebycheff coordinates}
This section contains results of computation of sl$_2$-valued ZCR's for
reduced GMC equations in Tchebycheff coordinates.
The method used is taken from \cite{M} and shortly explained at the end
of this section.

Fix an arbitrary system of \emph{Tchebycheff coordinates} $x$, $y$.
Then we have
\begin{eqnarray*}
g &=& dx^2 + 2 \cos f\,dx\,dy + dy^2, \\
b &=& b_{11}\,dx^2 + 2 b_{12}\,dx\,dy + b_{22}\,dy^2.
\end{eqnarray*}
The Gauss--Mainardi--Codazzi equations \cite{S},~eq.~(74b,c), are
\begin{eqnarray}
f_{,xy} &=& \frac{b_{12}^2 - b_{11} b_{22}}{\sin f}, \nonumber
\\
b_{11,y} &=& b_{12,x} + \frac{b_{22} - b_{12} \cos f}{\sin f} f_{,x},
\label{gmc}
\\
b_{12,y} &=& b_{22,x} - \frac{b_{11} - b_{12} \cos f}{\sin f} f_{,y} \nonumber.
\end{eqnarray}

There always exists a nonparametric $\Sl$-valued zero-curvature
representation (see \cite{S},~eq.~(84)), derived from the Gauss--Weingarten
equations. In Tchebycheff coordinates we have
\begin{equation} \label{zcr0}
\begin{array}{l} \displaystyle
A = \frac\Ii2\pmatrix f_x
& \displaystyle
\frac{\E^{\Ii f} b_{11} - b_{12}}{\sin f}
\\ \displaystyle
\frac{\E^{-\Ii f} b_{11} - b_{12}}{\sin f}
&
-f_x\quad
\endpmatrix,\\[10\jot]
\displaystyle
B = \frac\Ii2\pmatrix 0
& \displaystyle
\frac{\E^{\Ii f} b_{12} - b_{22}}{\sin f}
\\ \displaystyle
\frac{\E^{-\Ii f} b_{12} - b_{22}}{\sin f}
&
0
\endpmatrix.
\end{array}
\end{equation}

It is well known that this zero-curvature representation does not belong
to any $1$-parametric family.
However, imposing one additional relation between the unknowns
$x,y,f,b_{11},b_{12},b_{22}$ we obtain four distinct classes of surfaces
admitting a 1-parametric ZCR in the above sense.
The following proposition may be proved by straightforward computation.
We obtained it by methods of \cite{M}.

\begin{proposition}
\label{Prop1}
Let $X_1(x), X_2(x), Y_1(y), Y_2(y)$ be arbitrary functions and
$Z$ be a constant.
Denote
\begin{equation}
\label{L}
\begin{array}{l}
L :=  (b_{11} b_{22} - b_{12}^2) Z
 + Y_1 b_{11} + X_1 b_{22}
\\[2\jot]\displaystyle\qquad
 +\ [(X_2 - Y_2) \sin f - (X_1 + Y_1) \cos f] b_{12}
\\[2\jot]\displaystyle\qquad
 -\ (X_1 Y_1 + X_2 Y_2) \sin^2 f
 + (X_1 Y_2 - X_2 Y_1) \sin f \cos f.
\end{array}
\end{equation}
If $L = 0$, then the matrices
\[
\begin{array}{@{}l} \displaystyle
A = \frac12\pmatrix
\Ii f_x
& \displaystyle
\Ii \frac{\E^{\Ii f} b_{11} - b_{12}}{\sin f} + X_1 - \Ii X_2
\\[3\jot] \displaystyle
\Ii \frac{\E^{-\Ii f} b_{11} - b_{12}}{(Z + 1)\sin f} - \frac{X_1 + \Ii
X_2}{Z + 1}
& -\Ii f_x
\endpmatrix,
\\[10\jot]
B = \displaystyle\frac12\pmatrix
0
& \displaystyle
\Ii \frac{\E^{\Ii f} b_{12} - b_{22}}{\sin f} + (Y_1 - \Ii Y_2) \E^{\Ii f}
\\ \displaystyle
\Ii \frac{\E^{-\Ii f} b_{12} - b_{22}}{(Z + 1)\sin f}
 - \frac{Y_1 + \Ii Y_2}{Z + 1} \E^{-\Ii f}
&
0
\endpmatrix
\end{array}
\]
give a zero-curvature representation for equation~\eqref{gmc}.
\end{proposition}

Assume now that the functions $X_1,X_2,Y_1,Y_2,Z$ depend on
an auxiliary variable~$t$.
To be a possible parameter of the ZCR,
$t$ must not be a parameter of the equation~\eqref{gmc} itself.
In other words, the condition $L = 0$ must not depend on $t$.

If ${\partial L}/{\partial t} = 0$, then one easily derives from \eqref{L}
that $X_1 = X_2 = Y_1 = Y_2 = Z = 0$ and the ZCR is the same as~\eqref{zcr0},
without any parameter.

If ${\partial L}/{\partial t} - L/t = 0$, then $L(t) = t L(1)$, and the
condition $L(t) = 0$ does not depend on $t$ either.
Combining with \eqref{L}, we arrive at the classification given below.
We use the notation
\[
K = \frac{b_{11} b_{22} - b_{12}^2}{\sin^2 f}
\]
for the Gauss curvature
and
\[
H = \frac{b_{11} - 2 b_{12} \cos f + b_{22}}{2\sin^2 f}
\]
for the mean curvature.
Functions $X_i,Y_i$ are as in the proposition above.

\bigskip
\noindent\emph{Case 1.} $X_1 = Y_1 \ne 0$.\
This class corresponds to \emph{linear Weingarten surfaces} determined by a
linear relation
\[
K + 2 \kappa H + \lambda = 0
\]
between the two curvatures.
Here $\kappa, \lambda$ are arbitrary constants.

Our $t$-parametrized zero-curvature representation is
\[
\begin{array}{@{}l@{}} \displaystyle
A = \frac12\pmatrix
\Ii f_x&
\displaystyle
\Ii\,\frac{\E^{\Ii f} b_{11} - b_{12}}{\sin f} + \Delta^-(\kappa,\lambda,t)
\\[3\jot] \displaystyle
\Ii\,\frac{\E^{-\Ii f} b_{11} - b_{12}}{(t + 1)\sin f} -
\frac{\Delta^+(\kappa,\lambda,t)}{t + 1}& -\Ii f_x
\endpmatrix,
\\[12\jot]
B = \displaystyle\frac12
\pmatrix
\hskip -5mm  0&
\hskip -5mm
\displaystyle\Ii\,\frac{\E^{\Ii f} b_{12} - b_{22}}{\sin f}
 + \Delta^-(\kappa,\lambda,t)\E^{\Ii f}
\\[3\jot]
\displaystyle
\Ii\,\frac{\E^{-\Ii f} b_{12} - b_{22}}{(t + 1)\sin f}
 - \frac{\Delta^+(\kappa,\lambda,t)}{t + 1}\E^{-\Ii f} \hskip -5mm
& 0 \hskip -5mm
\endpmatrix,
\end{array}
\]
where $\Delta^\pm(\kappa,\lambda,t)=\kappa t\pm\sqrt{\kappa^2 t^2 +
\lambda t}$.

\bigskip
\noindent\emph{Case 2. } $X_1 = Y_1 = 0$.
If
\[
K + 2 \kappa \frac{b_{12}}{\sin f} - \lambda  = 0,
\]
where $\kappa, \lambda$ are constants, then there is
a $t$-parametrized zero-curvature representation
\[
\begin{array}{@{}l} \displaystyle
A = \frac\Ii2\pmatrix
f_x&
\displaystyle\frac{\E^{\Ii f} b_{11} - b_{12}}{\sin f}
- \Delta^+(\kappa,\lambda,t)
\\[4\jot] \displaystyle
\frac{\E^{-\Ii f} b_{11} - b_{12}}{( t + 1)\sin f}
 - \frac{\Delta^+(\kappa,\lambda,t)}{ t + 1}
&-f_x
\endpmatrix,
\\[12\jot]
B = \displaystyle\frac\Ii2\pmatrix
\hskip -5mm 0& \hskip -5mm
\displaystyle\frac{\E^{\Ii f} b_{12} - b_{22}}{\sin f}
 + \Delta^-(\kappa,\lambda,t)\E^{\Ii f}
\\[4\jot] \displaystyle
\frac{\E^{-\Ii f} b_{12} - b_{22}}{( t + 1)\sin f}
 + \frac{\Delta^-(\kappa,\lambda,t)}{ t + 1} \E^{-\Ii f} \hskip -5mm
& 0 \hskip -5mm
\endpmatrix.
\end{array}
\]

\bigskip
\noindent\emph{Case 3a. } $X_1 = 0$, $Y_1 \ne 0$.
If
\[
K \sin^2 f - Y_1 b_{11} + (Y_1 \cos f + Y_2 \sin f) b_{12} = 0,
\]
then
\[
\begin{array}{@{}l} \displaystyle
A = \frac\Ii2\pmatrix
f_x
& \displaystyle
\frac{\E^{\Ii f} b_{11} - b_{12}}{\sin f}
\\[3\jot] \displaystyle
\frac{\E^{-\Ii f} b_{11} - b_{12}}{( t + 1)\sin f}
& -f_x\quad
\endpmatrix,
\\[10\jot]
B = \displaystyle\frac\Ii2\pmatrix
\hskip -5mm 0
& \hskip -5mm \displaystyle
\frac{\E^{\Ii f} b_{12} - b_{22}}{\sin f} + (\Ii Y_1 + Y_2) t \E^{\Ii f}
\\ \displaystyle
\frac{\E^{-\Ii f} b_{12} - b_{22}}{( t + 1)\sin f}
 - \frac{\Ii Y_1 - Y_2}{ t + 1} t \E^{-\Ii f}\hskip -5mm
&
0 \hskip -5mm
\endpmatrix.
\end{array}
\]
is a $t$-parametrized zero-curvature representation.

\bigskip
\noindent\emph{Case 3b. } $X_1 \ne 0$, $Y_1 = 0$:
The relation is
\[
K\sin^2 f + (X_1 \cos f + X_2 \sin f) b_{12} - X_1 b_{22} = 0
\]
and the $t$-parametrized zero-curvature representation is
\[
\begin{array}{@{}l} \displaystyle
A = \frac\Ii2\pmatrix
f_x
& \displaystyle
\frac{\E^{\Ii f} b_{11} - b_{12}}{\sin f} + (\Ii X_1 - X_2) t
\\[3\jot] \displaystyle
\frac{\E^{-\Ii f} b_{11} - b_{12}}{( t + 1)\sin f}
 - \frac{\Ii X_1 + X_2}{ t + 1} t
& -f_x
\endpmatrix,
\\[10\jot]
B = \displaystyle\frac\Ii2\pmatrix
0
& \displaystyle
\frac{\E^{\Ii f} b_{12} - b_{22}}{\sin f}
\\ \displaystyle
\frac{\E^{-\Ii f} b_{12} - b_{22}}{( t + 1)\sin f}
&
0
\endpmatrix.
\end{array}
\]
Cases 3a and 3b transform one to another under
the $x \leftrightarrow y$ symmetry.

\bigskip
\noindent\emph{Case 4.} $0 \ne X_1 \ne Y_1 \ne 0$:
If
\[
K \sin^2 f
 - Y_1 b_{11}
 + (X_1 \E^{\Ii f} + Y_1 \E^{-\Ii f}) b_{12}
 - X_1 b_{22}
= 0,
\]
then
\[
\begin{array}{@{}l} \displaystyle
A = \frac\Ii2\pmatrix
f_x
& \displaystyle
\frac{\E^{\Ii f} b_{11} - b_{12}}{\sin f}
\\[3\jot] \displaystyle
\frac{\E^{-\Ii f} b_{11} - b_{12}}{( t + 1)\sin f}
 - \frac{2\Ii X_1 t}{ t + 1}
& -f_x
\endpmatrix,
\\[10\jot]
B = \displaystyle\frac\Ii2\pmatrix
0
& \displaystyle
\frac{\E^{\Ii f} b_{12} - b_{22}}{\sin f}
\\ \displaystyle
\frac{\E^{-\Ii f} b_{12} - b_{22}}{( t + 1)\sin f}
 - \frac{2 \Ii Y_1 t}{ t + 1} \E^{-\Ii f}
&
0
\endpmatrix.
\end{array}
\]
is a $t$-parametrized zero-curvature representation.

Case 4 is incompatible with the real geometry of $E^3$.
A formally real class is obtained after transformation
$f \mapsto \Ii f$, $x \mapsto \Ii x$, and $y \mapsto \Ii y$.
\bigskip

Two natural questions arise:
whether the parameter $t$ cannot be removed by gauge transformation
and whether there might be integrable cases outside this classification.
To give our answers, we first recall some facts from~\cite{M}.

In the simplest style, still applicable to the GMC equations, for a
system of PDE $\{F^l = 0\}_{l = 1}^N$ the {\it characteristic $N$-tuple}
of a ZCR $A\,dx + B\,dy$ consists of sl$_2$ matrices $C_l$,
$l = 1,\dots,N$, satisfying $A_{,y} - B_{,x} + [A,B] = C_l F^l$.
Then the gauge-equivalent ZCR with respect to a gauge matrix $S$ has
$\{S C_l S^{-1}\}_{l = 1}^N$ as its characteristic $N$-tuple.
Incidentally, in all four cases the characteristic triple is one and the
same:
$$
\def\arraystretch{1.7}
\frac \Ii 2
\left(\begin{array}{cr}
  1 & 0 \\ 0 & -1
\end{array}\right), \quad
\frac \Ii {2 \sin f}
\left(\begin{array}{cr}
  0 & \E^{\Ii f} \\  \displaystyle\frac{\E^{-\Ii f}}{t+1} & 0
\end{array}\right), \quad
\frac {-\Ii} {2 \sin f}
\left(\begin{array}{cr}
  0 & 1 \\  \displaystyle\frac1{t+1} & 0
\end{array}\right).
$$
As one obviously cannot remove $t$ by conjugation, $t$ is a true parameter.

To give at least partial answer to our second question, we explain the
assumptions we used in derivation of Proposition~\ref{Prop1}.
By~\cite{M}, the characteristic triple $C = \{C_1,C_2,C_3\}$ also satisfies
\begin{equation}\label{sys}
\widehat{R^\ast}(C) = 0,
\end{equation}
where $R^\ast$ is the formal adjoint to the operator of universal
linearization, and $\widehat{R^\ast}$ is obtained from $R^\ast$ by
replacement of total derivatives $D_x, D_y$ with ``covariant total
derivatives'' $D_x - \ad_A$, $D_y - \ad_B$ (or, equivalently, by lifting the
operator $R^\ast$ to the corresponding covering space by means of the
connection defining the covering).
The procedure then consists in solving the equation
$\widehat{R^\ast}(\bar C) = 0$
on unknowns $A,B,\bar C$, where $\bar C$ is a normal form of $C$ with
respect to conjugation.

For the above ZCR~\eqref{zcr0}, the characteristic triple
$C = \{C_1,C_2,C_3\}$ is
\begin{equation} \label{chi0}
\frac\Ii2
\left(\begin{array}{cr}
1 & 0 \\
0 & -1
\end{array}\right),
\quad
\frac\Ii{2 \sin f}\left(
\begin{array}{cc}
0 & \E^{\Ii f} \\
\E^{-\Ii f} & 0
\end{array}\right),
\quad
-\frac\Ii{2\sin f}\left(
\begin{array}{cc}
0 & 1 \\ 1 & 0
\end{array}\right).
\end{equation}

In \eqref{chi0}, $C_1$ is in Jordan normal form and is
diagonal.
Then, if the zero-curvature representation $(A,B)$ belongs to a 1-parameter
family, then the Jordan form for
$C_1$ will be diagonal for adjacent members of the family as well.
Similarly, since $C_2,C_3$ are both non-diagonal, they will be so for the
adjacent members.
By conjugation with an appropriate diagonal matrix leaving $C_1$ unchanged,
we may set one of the non-diagonal terms of $C_2$ or $C_3$ to~1.
Consequently, during our computation we may assume the characteristic element
to be of the form
$$
C_1 = \left(\begin{array}{cr}
r & 0 \\ 0 & -r
\end{array}\right),
\quad
C_2 = \left(\begin{array}{cr}
p_1 & p_2 \\ p_3 & -p_1
\end{array}\right),
\quad
C_3 = \left(\begin{array}{cr}
q_1 & 1\;\; \\ q_3 & -q_1
\end{array}\right).
$$

The functions $r,p_1,p_2,p_3,q_1,q_3$ together with the entries
of the sl$_2$-matrices
$$
A = \left(\begin{array}{cr}
a_1 & a_2 \\ a_3 & -a_1
\end{array}\right),
\quad
B = \left(\begin{array}{cr}
b_1 & b_2 \\ b_3 & -b_1
\end{array}\right)
$$
are twelve unknowns that are to be determined
along with the yet unknown dependence of $b_{22}$ on $x,y,f,b_{11},b_{12}$.

The determining system~\eqref{sys} is rather complex (typeset here, it would
fill approximately four pages) and manageable only with
the aid of computer algebra.
To solve it, we demanded that the expressions
$a_1,b_1$ depend on $x,y,f,f_x,f_y$, the expressions $a_2,a_3,b_2,b_3$ depend on
$x,y,f,b_{11}$, $b_{12}$ (cf.~eq.~\eqref{zcr0}),
and the expressions
$r,p_1,p_2,p_3,q_1,q_3$ depend on $x,y,f$ (cf.~\eqref{chi0}).
Finally, we demanded that the function $L$ in the constraint $L = 0$
actually depends on $b_{11},b_{12},b_{22}$ and is nonlinear in these variables.
Under assumptions stated, the result in Proposition~\ref{Prop1} is exhaustive.

\section{Geodesic coordinates}

Assume now that the immersed surface is endowed with the geodesic
coordinates, so that the metric is $g = dx^2 + f\,dy^2$, $f > 0$.
As before, the second fundamental form is
$b = b_{11}\,dx^2 + 2b_{12}\,dx\,dy + b_{22}\,dy^2$.
Then the Gauss--Mainardi--Codazzi equations are
\begin{eqnarray}
f_{,xx} &=& \frac{f_{,x}^2}{2f} + 2(b_{12}^2 - b_{11} b_{22}), \nonumber
\\
b_{11,y} &=& b_{12,x} + \frac{f_{,x}}{2f} b_{12}, \label{gmc geo}
\\
b_{12,y} &=& b_{22,x} + \frac{f_{,y}}{2f} b_{12}
 - \frac{f_{,x}}{2f} (b_{22} + f b_{11}) \nonumber.
\end{eqnarray}
The nonparametric zero-curvature representation is $A\,dx + B\,dy$ with
\begin{equation} \label{zcr0 geo}
\begin{array}{l} \displaystyle
A = -\frac12\left(
\begin{array}{cc}
0 & \displaystyle \frac{\Ii b_{12}}{\sqrt{f}} + b_{11}
\\
\displaystyle \frac{\Ii b_{12}}{\sqrt{f}} - b_{11} & 0\quad
\end{array}\right),
\\[10\jot] \displaystyle
B = -\frac12\left(
\begin{array}{cc}\displaystyle \frac{\Ii f_x}{2\sqrt{f}}
& \displaystyle \frac{\Ii b_{22}}{\sqrt{f}} + b_{12}
\\ \displaystyle \frac{\Ii b_{22}}{\sqrt{f}} - b_{12}
& \displaystyle -\frac{\Ii f_x}{2\sqrt{f}}
\end{array}\right).
\end{array}
\end{equation}

\begin{proposition}
\label{Prop2}
Let $Y_1(y), Y_2(y), Y_3(y), Y_4(y)$ be arbitrary functions
and $Z$ be a constant.
Use the notation
\begin{equation} \label{L geo}
\begin{array}{l}\displaystyle
L :=  \frac{1 - Z}{Z \sqrt{\!f}}  (b_{11} b_{22} - b_{12}^2)
  - \left(Y_1 \sqrt{\!f} + x \frac{\partial Y_2}{\partial y} + Y_4 \right)
    b_{11}
\\[4\jot]\displaystyle\qquad
  + \left(x \frac{\partial Y_1}{\partial y} + Y_3 \right)
    \frac 1{\sqrt{\!f}} \, b_{12}
  - \frac {Y_1}{\sqrt{\!f}} \, b_{22}
\\[4\jot]\displaystyle\qquad
  - Y_1 Y_4 + Y_2 Y_3
  - x Y_1 \frac{\partial Y_2}{\partial y}
  + x \frac{\partial Y_1}{\partial y} Y_2
  - (Y_1{}^2 + Y_2{}^2) \sqrt{\!f}.
\end{array}
\end{equation}
If $L = 0$, then the matrices
\[
\begin{array}{l@{}} \displaystyle
A = -\frac12\pmatrix
0 &
\displaystyle \left(\frac{\Ii b_{12}}{\sqrt{\!f}} + b_{11} + \Delta^-\right) Z
\\
\displaystyle \frac{\Ii b_{12}}{\sqrt{\!f}} - b_{11} - \Delta^+ & 0\quad
\endpmatrix,
\\[10\jot] \displaystyle
B = -\frac12 \\[.2in]
\times \pmatrix \hskip-2cm\displaystyle  \frac{\Ii f_x}{2\sqrt{\!f}}
& \hskip-2.4cm
 \biggl( \displaystyle\frac{\Ii b_{22}}{\sqrt{\!f}} + b_{12}
 + x \frac{\partial \Delta^+}{\partial y} + \Ii\Delta^+ \sqrt{\!f}
 + Y_3 + \Ii Y_4\biggr) Z
\\[2ex]
\displaystyle \frac{\Ii b_{22}}{\sqrt{\!f}} - b_{12}
 - x \frac{\partial \Delta^-}{\partial y}  + \Ii \Delta^- \sqrt{\!f}
 - Y_3 + \Ii Y_4 \hskip-2.2cm
& \displaystyle -\frac{\Ii f_x}{2\sqrt{\!f}} \hskip -1.5cm
\endpmatrix,
\end{array}
\]
where $\Delta^\pm=Y_1 \pm \Ii Y_2$,
form a zero-curvature representation for equation~\eqref{gmc geo}.
\end{proposition}

Denoting
\[
K = \frac{b_{11} b_{22} - b_{12}^2}{f},
\qquad
H = \frac12 \left(b_{11} + \frac{b_{22}}{f} \right)
\]
the Gauss and mean curvature, respectively,
we get the following five integrable classes.

\bigskip
\noindent\emph{Case 1. } Linear Weingarten surfaces
\[
K + \alpha H + \beta = 0,
\]
where $\alpha,\beta$ are arbitrary constants. Then
\begin{displaymath}
\begin{array}{@{}l}
A = \displaystyle-\frac12\pmatrix
0
&\displaystyle
(t + 1) \left(\frac {\Ii b_{12}}{\sqrt f} + b_{11}\right)
 + \square^-
\\\displaystyle
\frac {\Ii b_{12}}{\sqrt f} - b_{11}
 - \frac {\square^+}{t + 1}
&0
\endpmatrix,
\\\noalign{\medskip}\displaystyle
B = -\frac12 \pmatrix
\hskip -5mm \displaystyle\frac {\Ii f_{x}}{2 \sqrt f}
& \hskip -5mm \displaystyle
(t + 1)\left(\frac {\Ii b_{22}}{\sqrt f} + b_{12}\right)
 + \square^-\,\Ii\sqrt f
\\[4mm]\displaystyle
\frac {\Ii b_{22}}{\sqrt f} - b_{12}
 + \frac{\square^+}{t + 1}\,\Ii\sqrt f \hskip -5mm
&\displaystyle
-\frac {\Ii f_{x}}{2 \sqrt f} \hskip -5mm
\endpmatrix,
\end{array}
\end{displaymath}
where $\square^\pm=\square^\pm(\alpha,\beta,t)=
\alpha t \pm \sqrt{\alpha^2 t^2 - \beta t - \beta t^2}$.

\bigskip
\noindent\emph{Case 2. }
The class
\[
K + Y \frac{b_{11}}{\sqrt f} + \gamma = 0,
\]
where $Y$ is an arbitrary function of $y$ and $\gamma$ is a constant. In
this case
\begin{displaymath}
\def\arraystretch{2.8}
\begin{array}{@{}l}\displaystyle
A = -\frac12\pmatrix
0 & \displaystyle
(t+1)\left(\frac {\Ii b_{12}}{\sqrt f} + b_{11}
 + \Ii \sqrt {\frac {\gamma t}{t+1}}\right)
\\\displaystyle
\frac {\Ii b_{12}}{\sqrt f} - b_{11}
 + \Ii \sqrt {\frac {\gamma t}{t+1}} &
0
\endpmatrix,
\\\displaystyle
B = -\frac12\pmatrix\displaystyle
\hskip -7mm \frac {\Ii f_{x}}{2\sqrt {f}}
&\hskip -7mm \displaystyle
(t+1)
\left(\frac {\Ii b_{22}}{\sqrt f} + b_{12} + \bar{Y}
 - \sqrt {\frac {\gamma f t}{t+1}}\right)
\\\displaystyle
\frac {\Ii b_{22}}{\sqrt f} - b_{12} + \bar{Y}
 + \sqrt {\frac {\gamma f t}{t+1}} \hskip -7mm
&\displaystyle
-\frac {\Ii f_{x}}{2\sqrt {f}} \hskip -7mm
\endpmatrix,
\end{array}
\end{displaymath}
where $\bar{Y}={\Ii\,t Y}/{(t+1)}$.

\bigskip
\noindent\emph{Case 3. }
The class
\[
K + Y_1 \frac{b_{11}}{\sqrt {f}} + Y_2 \frac{b_{12}}{f} = 0,
\]
where $Y_1,Y_2$ are arbitrary functions of $y$. Then
\begin{displaymath}
\def\arraystretch{2.5}
\begin{array}{@{}l}\displaystyle
A = -\frac12\pmatrix
0
&\displaystyle
(t + 1)\left(\frac{\Ii b_{12}}{\sqrt f} + b_{11}\right)
\\\displaystyle
\frac{\Ii b_{12}}{\sqrt f} - b_{11}
&
0
\endpmatrix,
\\\noalign{\medskip}\displaystyle
B = -\frac12\pmatrix
\hskip -5mm \displaystyle \frac {\Ii f_{x}}{2 \sqrt {f}}
&\hskip -5mm \displaystyle
(t + 1) \left(\frac {\Ii b_{22}}{\sqrt f} + b_{12}\right)
 + (\Ii Y_1 - Y_2)t
\\\displaystyle
\frac{\Ii b_{22}}{\sqrt f} - b_{12} + \frac {(\Ii Y_1 + Y_2) t}{t+1}
\hskip -5mm
&\displaystyle
-\frac {\Ii f_{x}}{2 \sqrt {f}}\hskip -5mm
\endpmatrix.
\end{array}
\end{displaymath}

\bigskip
\noindent\emph{Case 4. } The class
\[
K + Y_1 H - \frac12 Y \left(\frac{b_{12}}{f} + \Ii \frac{b_{11}}{\sqrt f}
\right) = 0,
\qquad
Y = x\frac{\partial Y_1}{\partial y} + Y_2,
\]
where $Y_1,Y_2$ are arbitrary functions of $y$. Then
\begin{displaymath}
\def\arraystretch{2.5}\begin{array}{l}\displaystyle
A = -\frac12\pmatrix
0 &\displaystyle
(t + 1)\left(\frac{\Ii b_{12}}{\sqrt f} + b_{11}\right) + Y_1 t
\\\displaystyle
\frac{\Ii b_{12}}{\sqrt f} - b_{11}
& 0
\endpmatrix,
\\\noalign{\medskip}\displaystyle
B = -\frac12\pmatrix\displaystyle
\frac {\Ii f_{x}}{2 \sqrt f}
& \displaystyle
(t + 1)\left(\frac {\Ii b_{22}}{\sqrt f} + b_{12}\right)
 + t Y + \Ii Y_1 t\sqrt {f}
\\\displaystyle
\frac {\Ii b_{22}}{\sqrt f} - b_{12}
&\displaystyle
-\frac {\Ii f_{x}}{2 \sqrt f}
\endpmatrix.
\end{array}
\end{displaymath}

Case 4 is incompatible with the real geometry of $E^3$, but
nevertheless we may turn it into a formally real class by allowing
$f$ to be negative.
\section*{Acknowledgements}
The authors are grateful to J.~Cie\'sli\'nski~J. and A.~Sym for calling
their attention to the problem. The send author would also like to thank
A.~Bobenko and E.~Ferapontov for helpful criticism.


\begin{thebibliography}{99}
\bibitem{B-N}
Barbashov~B.~M. and Nesterenko~V.~V.:
Geometricheskij analiz nelinejnykh uravnenij v teorii relyativistskoj struny,
[Geometrical analysis of equations in the theory of relativistic string]
\emph{Fiz. Elem. Chast. At. Yadra} \textbf{15}, no.~5 (1984) 1032--1072
(in Russian).

\bibitem{B1}
Bobenko~A.~I.:
Integriruemye poverkhnosti, [Integrable surfaces]
\emph{Funkc. Anal. Prilozh.} \textbf{24}, no.~3 (1990) 68--69 (in Russian).

\bibitem{B2}
Bobenko~A.~I.:
Surfaces in terms of 2 by 2 matrices. Old and new
integrable cases,  in: A.P. Fordy and J.C. Wood, eds.,
\emph{Harmonic Maps and Integrable Systems}
(Vieweg, Braunschweig, 1994) 83--127.

\bibitem{C}
Cie\'sli\'nski~J.:
A generalized formula for integrable classes of surfaces
in Lie algebras, \emph{J. Math. Phys.} \textbf{38} (1997) 4255--4272.

\bibitem{C-G-S}
Cie\'sli\'nski~J., Goldstein~P., and Sym~A.: Isothermic surfaces in $E^3$
as soliton surfaces, \emph{Phys. Lett. A} \textbf{205} (1995) 37--43.

\bibitem{K}
Krasil{\cprime}shchik~I.~S.:
Some new cohomological invariants for nonlinear differential equations,
\emph{Differential Geom. Appl.} \textbf{2} (1992), 307--350.

\bibitem{K1}
Krasil{\cprime}shchik~I.~S.: Notes on coverings and B\"acklund
transformations, Preprint ESI \textbf{260}, Wien,1995 (http://www.esi.ac.at).

\bibitem{K-L-V}
Krasil{\cprime}shchik~I.~S., Lychagin~V.~V., and Vinogradov~A.~M.:
\emph{Geometry of jet spaces and nonlinear partial differential equations},
Gordon and Breach, New York, 1986.

\bibitem{K-V}
Krasil{\cprime}shchik~I.S. and Vinogradov~A.~M.:
\emph{Nonlocal trends in the geometry of differential equations:
{S}ymmetries, conservation laws, and {B\"a}cklund transformations},
Acta Appl. Math. \textbf{15} (1989), 161--209.

\bibitem{L-S}
Levi~D. and Sym~A.: Integrable systems describing surfaces of
non-constant curvature, \emph{Phys. Lett. A} \textbf{149} (1990) 381--387.

\bibitem{M}
Marvan~M.:
On zero curvature representations of partial differential equations,
in: \emph{Differential Geometry and Its Applications}, Proc. Conf. Opava,
Czechoslovakia, Aug. 24--28, 1992 (Silesian University, Opava, 1993)
103--122. Electronic version available from ELibEMS.

\bibitem{M1}
Marvan~M.:
Another look on recursion operators, in: Differential Geometry and
Applications, Proc. Conf. Brno, 1995 (Masaryk University, Brno, 1996)
393--402.

\bibitem{M-S}
Melko~M. and Sterling~I.:
Integrable systems, harmonic maps and the classical theory of surfaces,  in:
A.P. Fordy and J.C. Wood, eds., \emph{Harmonic Maps and Integrable Systems}
(Vieweg, Braunschweig, 1994) 129--144.

\bibitem{S}
Sym~A.:
Soliton surfaces and their applications (soliton geometry from spectral
problems), in: R. Martini, ed., \emph{Geometric Aspects of the Einstein
Equations and Integrable Systems}, Proc. Conf. Scheveningen, The
Netherlands, August 26--31, 1984, Lecture Notes in Physics 239 (Springer,
Berlin et al., 1985) 154--231.

\bibitem{Te}
K. Tenenblat, {\it Transformations of Manifolds and Applications to
Differential Equations} (Addison Wesley Longman, Harlow, 1998).

\bibitem{Ts}
Tsujishita~T.:
Homological method of computing invariants of systems of
differential equations, {\it Diff. Geom. Appl.} {\bf 3} (1991) 3--34.

\bibitem{W}
Wu~H.:
Weingarten surfaces and nonlinear partial differential equations,
\emph{Ann. Global An. Geom.} \textbf{11} (1993) 49--64.

\end{thebibliography}
\end{document}